\documentclass[a4paper]{llncs}
\usepackage[english]{babel}
\usepackage[T1]{fontenc}
\usepackage{amsmath}
\usepackage{amsfonts}
\usepackage{ulem}%to use \sout for stroke
\usepackage{amssymb}
\usepackage{epsfig}%For figures
\usepackage{graphicx}
\usepackage{array}
\usepackage{pxfonts}
\usepackage{setspace}
\usepackage{algorithm,algorithmic}
\newtheorem{define}{Definition}

\title{Data Delivery by Mobile Agents with Energy Constraints over a fixed path}
\author{Aristotelis Giannakos\inst{1,2} \and Mhand Hifi\inst{2} \and Gregory Karagiorgos\inst{3} }

\institute{PSL Research University, Université Paris-Dauphine, LAMSADE, CNRS UMR 7243, Paris, France \\ \email{\bf giannako@lamsade.dauphine.fr}
\and
EPROAD-EA 4669, Université de Picardie Jules Vernes, 7 rue du Moulin Neuf, 80000 Amiens, France \\ \email{\{\bf aristotelis.giannakos, mhand.hifi\}@u-picardie.fr}
\and
Department of Computer Engineering, Technological Educational Insitute of Peloponnese, Kladas, 23100 Sparta, Greee \\ \email{\bf greg@teikal.gr}
}

\begin{document}

\maketitle

\date{}
%\doublespace

\begin{abstract}
We consider $k$ mobile agents of limited energy that are initially located at vertices of an edge-weighted graph $G$ and have to collectively deliver data from a source vertex $s$ to a target vertex $t$. The data are to be collected by an agent reaching $s$, who can carry and then hand them over another agent etc., until some agent with the data reaches $t$.
The data can be carried only over a fixed $s-t$ path of $G$; each agent has an initial energy budget and each time it passes an edge, it consumes the edge's weight in energy units and stalls if its energy is not anymore sufficient to move. The main result of this paper is a 3-approximation polynomial time algorithm for the data delivery problem over a fixed $s-t$ path in the graph, for identical initial energy budgets and at most one allowed data hand-over per agent. 


\end{abstract}

\section{Introduction}

The use of commodity mobile robots is increasingly replacing human labor in various coordinated tasks. One of the fundamental problems that need to be addressed for further advancing in this process, is the delivery of data from one location to another.
With the density of energy storage being still perhaps the main technological bottleneck in mobile robots
development, the need for energy-efficient operations is a key factor that makes tasks such as data-delivery far from trivial.

In this paper we deal with the problem of delivering data from a source $s$ to a target $t$, by using a set
\{1, 2, \ldots, k\} of mobile agents (robots) that lie initially at position $q_1, \ldots, q_k,$ respectively.
The data  are picked up by some agent that arrives to $s,$ and carried for  some distance  before  being
handed  over to another agent etc, until some agent with the data reach $t$. The space where the agents move 
is modeled by an edge-weighted graph (whose vertices include $s, t, q_1, \ldots, q_k$).
A mobile agent can pass from a location (represented by a vertex) $a$ to a neighboring location $b$, only if it has
energy greater than or equal to the amount it will consume to traverse the edge with endpoints $a$ and $b$,
which is the weight of that edge. Initially, each robot $i$ has energy $r_i$.
We further assume that $(i)$ all agents are of the same type and in particular, all the $r_{is}$ are equal,
$(ii)$ each agent is simple-built so that it can be over-handed data at most once and $(iii)$ data can travel only
on a pre-specified $s-t$ "secure" path $\cal{P}$ of the graph \cite{CDMPW}.
In this setting, we seek a schedule for agents of minimum initial energy $r$, that deliver the
data from $s$ to $t$ following $\cal{P}$, where each agent used is being handed data only once.   
We refer to this problem as {\sc Min-Range Uniform Data Delivery over a Path (min-RIPDD)}; it 
is shown NP-hard in \cite{CDMPW}.
In this paper, we prove a 3-approximation polynomial time algorithm for {\sc min-RIDD}; to our knowledge this is the first approximation algorithm for this problem.

{\bf Related work:} Energy-aware task performing by mobile agents is an emerging research area, in particular with respect to algorithm-theoretical problematics; see for instance \cite{RV} where self-deployment of agents is investigated in this context. On the contrary, problems dealing with minimizing of the total distance run by unlimited (ie always sufficient) energy-powered robots are well-studied \cite{AH,AG,ACCLPV}.
 
Anaya et al. \cite{ACCLPV} tackle the so-called {\it convergecast} problem, which is a specific variant of aggregation of data initially partitioned into pieces detained by different agents: the latter move over an edge-weighted graph, consuming energy (within the limit of their initial identical budgets) equal to the weight of an edge each time they transverse it; each time they meet, they exchange data, the objective being to have at some point at least one agent disposing all data. This problem is polynomial for path graphs, but becomes NP-hard for trees.

The first results on data delivery for energy-aware mobile agents are shown by Chalopin et al. in \cite{CDMPW}; they give a first formal setting of the problem, they characterize the complexity status for several variants of it and they prove a 2-approximation algorithm for the case of identical energy budgets. In \cite{CJMW} the authors they show the NP-completeness for the feasibility of data delivery when the graph is a path; they also prove a pseudo-quasipolynomial algorithm for deciding it.
Finally, in a recent paper \cite{BCDDGGLM} Batschi et al. show a polynomial algorithm for data delivery in the case where the agents have to return to their respective initial positions and the underlying graph is a tree. 

The paper is organized as follows. In Section 2 we give some definitions and notations necessary to introduce a formal setting for data delivery problems, which are presented in Section 3. Section 4 is devoted to the presentation of our results. Finally, the last Section outlines some interesting question for future work in this vein.

\section{Definitions and preliminaries}
Let $G(V,E)$ be a simple finite graph without loops with vertex set $V$ and edge set $E$ (denoted also by $V(G)$ and $E(G)$, respectively). We denote $|V|$ by $n$ and $|E|$ by $m$. An edge from $v$ to $u$ will be denoted by $vu$; if $G$ is non-directed, $vu$ and $uv$ will be used without distinction to denote the edge between $v$ and $u$. For $V'\subset V$, $\Gamma(V')=\{u|u\notin V' \text{ and } vu\in E\}$ and $N(V')=\Gamma(V')\cup V'$; the subgraph of $G$ induced by $V',$  ie $G'(V', E'=E\cap V'^2)$ will be denoted by $G[V']$. \\
We use the standard convention of denoting vectors of any kind by characters in boldface. Let $\displaystyle \mathbf{x}=\left(x_j\right)_{j=1}^l$ and $\displaystyle \mathbf{y}=\left(y_j\right)_{j=1}^l$ be vectors of $l$ components. Then: $(\mathbf{x}_{-j}, z)$ is obtained by replacing in $\mathbf{x}$ the component $x_j$ by $z$; we also write $\mathbf{x}\geq \mathbf{y}$ whenever $\forall j,1\leq j\leq l$ it is $x_j\geq y_j$.
\begin{define}
A data delivery model $\cal{D}$ is a tuple $\langle G, w, S, T, k, \mathbf{q}^0, \mathbf{r}^0 \rangle$ where\\
-$G$ is a non-directed simple graph,\\ 
-$w$ is a function $E(G)\cup \{(v,v)|v\in V(G)\}\rightarrow \mathbb{Z}_{\geq 0}$ with $w(v,v)=0$, \\
-$S, T\subset V(G), \: S\cap T=\emptyset$,\\ 
-$k$ is a positive integer $\leq n$ and \\
-$\mathbf{q}^0\in V(G)^k, \mathbf{r}^0\in \mathbb{N}^k$. 
\end{define}
\begin{define}\label{feasibility}
We say that a data delivery model ${\cal D}=\langle G, w, S, T, k, \mathbf{q}^0, \mathbf{r}^0 \rangle$ is {\it feasible} iff there is a feasibility suite, ie a suite  $\mathbf{p}^0,\ldots, \mathbf{p}^l$ such that\\
-$\mathbf{p}^0=(\mathbf{q}^0,\mathbf{r}^0)$\\
-for $1\leq i\leq l,$ it is $\mathbf{p}^i=\displaystyle \left( \mathbf{q}^{i}\in N(\mathbf{q}^{i-1}),\mathbf{r}^i=\left(r^{i-1}_j-w(q^{i-1}_j,q^i_j)\right)_{j=1}^k\right)$\\
-$ S\cup T$ belong to the same connected component in $G\left[\displaystyle\bigcup_{i=0}^l\mathbf{q}^i\right]$ and \\
-$\mathbf{r}^l\geq \mathbf{0}$\\
\end{define} 
A data delivery model can be interpreted intuitively as a situation where $k$ agents $1,\ldots, k$ that can move over the vertex set of a graph $G$ through its edges, are initially located at vertices $q^0_1, \ldots, q^0_k$, respectively. They have to collect data from the locations (vertices) of $S$ and transfer them to the ones of $T$. They have an initial energy budget $r^0_1,\ldots, r^0_k$, respectively. Whenever an agent traverses an edge $vu$, it consumes energy equal to $w(u,v)$; thus, any agent in order to be able to move over an edge, must have adequate energy and if it lies in a position $v$ having no more energy enough to traverse any edge $vu$, then it's left ``dead'' on $v$. Agents move in a synchronous mode by steps from a set of positions (vertices) to a neighboring and attainable one (given their respective energies), and they can exchange data with eachother whenever they are on a same position $v$ during some step.
%\begin{rem}

Non-weighted data delivery models, ie with $w(vu)=1 \: \forall vu\in E(G)$ can be used to model a relaxation of the above-described situation, namely the case where agents can also partially traverse edges (to the extend allowed by their available energy budgets) and meet along edges.  

%\end{rem} 
Indeed, to do so one has to transform $G, w$ in the following manner: first, contract all edges $uv $ for which $w(uv)=0$; then, replace any edge $uv$ with $w(uv)>0$ by a $u-v$ path of $w(uv)$ edges. Finally, put $w'(xy)=1$ for all edges $xy$ in the newly obtained graph $G'$.

\section{The data delivery problem}

In this Section we further refine the formal models introduced above.
\begin{define}
The {\sc data delivery (DD)} problem is to decide, given a $\cal{D}$, whether it is feasible.
\end{define}

In the sequel, we deal with the case where $S$ and $T$ are singletons (noted by $s$ and $t$, respectively). Under this assumption, we study variants of the data delivery problem defined below:
\begin{define}
Let ${\cal D}=\langle G, w, s, t, k, \mathbf{q}^0, \mathbf{r}^0 \rangle$ be a data delivery model with $0\leq r_j^0\leq 1,\: 1\leq j\leq k$. The {\sc min-range uniform data delivery (min-RUDD)} problem is to find the least $\rho\geq 1$ for which $\langle G, w, s, t, k, \mathbf{q}^0, \rho\mathbf{r}^0 \rangle$ becomes feasible.\\
The {\sc min-range identical data delivery (min-RIDD)} problem is the special case of 
{\sc min-RUDD} where $\mathbf{r}^0=\mathbf{1}$.
\end{define} 
A data delivery model is {\it feasible over a fixed $s-t\:$ path $\cal{P}$} iff it is feasible and $G\left[\displaystyle\bigcup_{i=0}^l\mathbf{q}^i\right]$ contains $\cal{P}$.
We define then, in a similar way as previously done, the following problems:
\begin{define} 
Given ${\cal D}=\langle G, w, s, t, k, \mathbf{q}^0, \mathbf{r}^0 \rangle$ and 
$\cal{P}$ an $s-t$path on $G$, the\\
{\sc data delivery over a path (PDD)} problem is to decide whether ${\cal D}$ is feasible over ${\cal P}$.\\
If $0\leq r_j^0\leq 1,\: 1\leq j\leq k$, the {\sc min-range uniform data delivery over a path\\ (min-RUPDD)} is to find the least $\rho\geq 1$ for which $\langle G, w, s, t, k, \mathbf{q}^0, \rho\mathbf{r}^0 \rangle$ becomes feasible over ${\cal P}$ (the special case of $\mathbf{r}^0=\mathbf{1}$ is called {\sc min-range identical data delivery over a path (min-RIPDD)} problem).
\end{define}

For the sake of completeness we mention the following important theorem:
%\section{Related work and some easy remarks}

\begin{theorem}{\rm \cite{CDMPW}}
\\
1.{\sc DD} and {\sc PDD} are $\mathbf{NP}-$complete.\\
2. {\sc min-RIPDD} is $\mathbf{NP}-$hard.
%2. Decide if {\sc min-RIPDD} has a solution of $\rho\leq 16$ is $\mathbf{NP}-$complete.
\end{theorem}

\section{Approximation algorithms}

The {\sc min-RIPDD} problem with at most one hand-over per agent may also be defined in a more straight-forward manner:\\
Let $G(V,E)$ be a weighted connected graph, $\cal{P}$ an $s-t$ path in $G$, $Q=\{q_1,\ldots, q_k\}\subset V$. Find paths $p_1,\ldots, p_k$ in $G$ such that: 
\begin{itemize}
\item[-] every $p_i$ has an endpoint in $Q$;  
\item[-] $\displaystyle \bigcup p_i$ contain $\cal{P}$ and every $p_i$ contains at most one connected component of $\cal{P}$;
\item[-] the length of the longest $p_i$ is minimum.
\end{itemize}
For this problem, the first intuitive idea may be the following naïve algorithm.
%%%%%%%%%%%%%%%%%%%%%%%%%%%%%%%%%%%%%%%%%%%%%%%%%%%%%%%%%
%\newpage

\subsection{A naïve algorithm for {\sc min-RIPDD}}
\begin{algorithm}
\caption{\bf{Greedy}$_{1}$}   %(assume that $\beta\!\mid\!\cal{P}$'s length $W$)\\
{\bf Input:} $G, {\cal P}, Q$ as above.\\
{\bf Output:} A feasible range $R$.
\begin{algorithmic}[1]
\STATE vertex spoint; robot crobot; set of agents $L$; array of integers indexed by $Q \:\: R[]$;\\
\STATE spoint$\leftarrow s$; $L\leftarrow \emptyset$; $R[]\leftarrow 0$;\\
\STATE {\bf while} $(\text{spoint}\neq t)$ {\bf do} 
\STATE \ \ \ $ $crobot$ \displaystyle\leftarrow \text{arg}\!\min_{q_i\not{in} L} \left\{\text{shortest path from } q_i \text{ to spoint}, R[\text{crobot}]\right\} $ ;

\STATE \ \ \ $p\leftarrow \text{shortest path from crobot to spoint}$;

\STATE \ \ \ $R[\text{crobot}]\leftarrow R[\text{crobot}]+p+1$;

\STATE \ \ \ $L\leftarrow L\cup \{\text{crobot}\}$;

\STATE \ \ \ $ $spoint$ \leftarrow$ the next vertex on $\cal{P}$;

\STATE {\bf endwhile};

\STATE {\bf return} $R\leftarrow \max_i R[i]$. 
\end{algorithmic}
\end{algorithm}

%\newpage

\begin{proposition}
Greedy$_{1}$ has approximation ratio $\geq \sqrt{n}$
\end{proposition}

{\bf Proof.} Consider the instance of the figure \ref{bigr1} below:
\begin{figure}[htb]
\fbox{\parbox[b]{.99\linewidth}{
\vskip 0.5cm
\begin{center}
\includegraphics[width=\columnwidth]{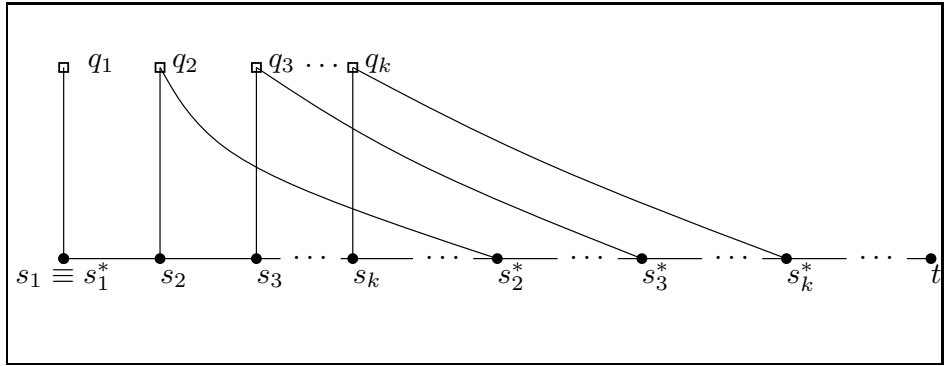}
\end{center}
\vskip 0.5cm}}
\caption{\protect\label{bigr1} A ''bad'' instance for Greedy$_{1}$. All non-valued edges have weight 1.}
\end{figure}

Assume that $k\leq \sqrt{n}$. Let $W$ be the total weight of $\cal{P}$. Greedy$_{1}$ will fetch agent $q_i$ to $s_i$, for $1\leq i \leq k$, leaving $q_k$ to walk through the remaining $W-k$ edges of $\cal{P}$ to reach $t$.

The optimal will take $q_i$ at $s^*_i$, with consecutive $s^*_i$s being distant $\frac{W}{k}$ to eachother. So the ratio would be, by this case, $\geq \frac{W-k+1}{\frac{W}{k}+1}$, which tends to $\sqrt{W}$ for $k=\sqrt{W}$ and large $W$.
   
Taking $W=n-k$ in an unweighted graph, would give\\ ratio $\displaystyle\geq \frac{(n+2+\sqrt{4n+1})(\sqrt{4n+1}-1)}{2n+1-\sqrt{4n+1}}$ which is greater than but tends asymptotically to $\sqrt{n}$.
\hfill$\qed$

\subsection{A greedy algorithm for {\sc min-RIPDD}}

Greedy$_{1}$ can be further generalized to the following algorithm:

\begin{algorithm}
\caption{ \bf{Greedy}$_{\beta}$}
{\bf Input:} Integer $\beta,\: 1<\beta\leq W$ where $W$ is $\cal{P}$'s length, and $G, {\cal P}, Q$ as above.\\
{\bf Output:} A feasible range $R$.
\begin{algorithmic}[1]
\STATE vertex spoint; robot crobot; set of robots $L$; array of integers indexed by $Q \:\: R[]$;\\
\STATE spoint$\leftarrow s$; $L\leftarrow \emptyset$; $R[]\leftarrow 0$;\\
\STATE {\bf while} $(\text{spoint}\neq t)$ {\bf do}

\STATE \ \ \ $ $crobot$ \displaystyle\leftarrow \text{arg}\!\min_{q_i\not{in} L} \left\{\text{shortest path from } q_i \text{ to spoint}, R[\text{crobot}]\right\} $ ;

\STATE \ \ \ $p\leftarrow \text{shortest path from crobot to spoint}$;

\STATE \ \ \ {\bf if} (spoint$=s$ and $(W\mod\beta)>0$) {\bf then} $b\leftarrow (W\mod\beta)$ {\bf else} $b\leftarrow \beta$;

\STATE \ \ \ $R[\text{crobot}]\leftarrow R[\text{crobot}]+p+b$;

\STATE \ \ \ $L\leftarrow L\cup \{\text{crobot}\}$;

\STATE \ \ \ $ $spoint$ \leftarrow$ the vertex reached after having walked $b$ on $\cal{P}$;

\STATE {\bf endwhile};\\
\STATE {\bf return} $R\leftarrow \max_i R[i]$.
\end{algorithmic}
\end{algorithm} 

%\newpage
\begin{proposition}
Greedy$_{\frac{W}{k}}$ has approximation ratio $\geq \lceil\frac{k}{2}\rceil$
\end{proposition}

{\bf Proof.} Consider the instance of the figure \ref{bigr2} below.
\begin{figure}[htb]
\fbox{\parbox[b]{.99\linewidth}{
\vskip 0.5cm
\begin{center}
\includegraphics[width=\columnwidth]{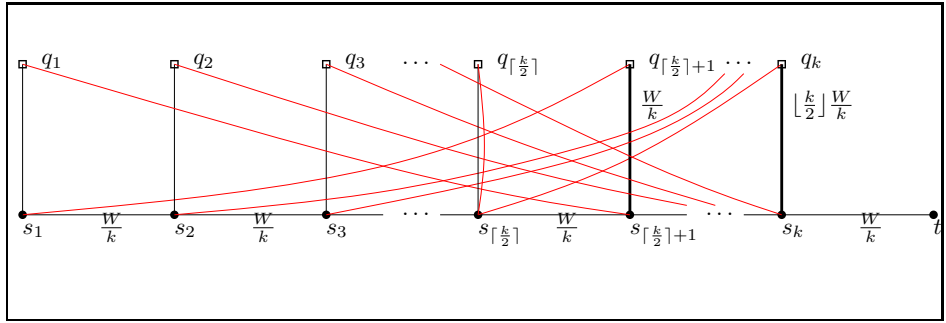}
\end{center}
\vskip 0.5cm}}
\caption{\protect\label{bigr2} A ''bad'' instance for Greedy$_{\frac{W}{k}}$. Red paths are taken by the optimal. Light non-valued edges have weight $\epsilon$, all others as depicted.}
\end{figure}

We assume that $W$ is the total weight of $\cal{P}$,$\: k|W$ and $k$ odd, which is slightly more complicated.

The optimal solution is to fetch agent $q_{\lceil\frac{k}{2}+1\rceil}$ to $s_1$, then for $2\leq i\leq \lfloor\frac{k}{2}\rfloor$ agent $q_{\lceil\frac{k}{2}\rceil+i}$ to selection point $s_i$, agent $q_{\lceil\frac{k}{2}\rceil}$ to $q_{\lceil\frac{k}{2}\rceil}$, and finally for $\lceil\frac{k}{2}\rceil+1\leq i \leq k$ agent $q_{i-\lceil\frac{k}{2}\rceil}$; this gives $R^*=\frac{W}{k}+\epsilon$.\\
Greedy$_{\frac{W}{k}}$ could fetch  $q_1$ to $s_1$ (since ties break arbitrarily). For all $2\leq i \leq \lceil\frac{k}{2}\rceil$, agent $q_{i-1}$ reaches $s_i$ and at this point $q_i$ could be chosen as a best candidate. Notice now that, from $s_{\lceil\frac{k}{2}\rceil+i}$ to $k$, for $1\leq i\leq \lfloor\frac{k}{2}\rfloor, \:\:q_i$ is the best candidate but with an augmenting overhead that achieves its maximum at the last part of the path; thus Greedy$_{\frac{W}{k}}$ would return $R=\lfloor\frac{k}{2}\rfloor\frac{W}{k}+\frac{W}{k}$ which bounds the ratio from below to $\lceil\frac{k}{2}\rceil$ as claimed.$\hfill\qed$

%\begin{rem}    
Notice that for $k=\sqrt{W}$, Greedy$_{\frac{W}{k}}$ does not seem to perform significantly better than Greedy$_{1}$. 
%\end{rem}

\subsection{A matching algorithm}
Given that the reason for Greedy$_{\beta}$'s not yieldig a constant ratio seems to be the short-sighted way a best candidate is chosen to be fetched at each selection point, a natural idea would be to improve this by assigning best candidates through a best matching between robots and selection points.

Indeed, one can form a complete weighted bipartite graph $B(Q,S_\beta,w)$ as follows:
\begin{itemize}
\item[-]$Q=\{q_1,\ldots, q_k\}$,
\item[-]$S_\beta=\{s_1\equiv s,\ldots, s_{\lfloor\frac{W}{\beta}\rfloor}|s_i\text{ points of }{\cal P}\}$ for some $\frac{W}{k}\leq \beta\leq W,$ with \\ $s_2\:(W\!\!\mod\!\beta)$-distant from $s_1$  and for $2<i\leq\lfloor\frac{W}{\beta}\rfloor, \: s_i\:\beta$-distant from $s_{i-1}$,
\item[-]$w(q_i, s_j)$ the length of a shortest path $q_i-s_j$ in $G$. 
\end{itemize}
Then a maximal matching $M_\beta$ of minimum maximum weight edge in $B(Q,S_\beta,w)$ can be computed easily at the cost of $O\left(k\frac{W}{\beta}\right)$ matchings, by repeatedly searching for a maximal matching in $B(Q,S_\beta,w)$ after removal of the maximum weight edges. 

Such a preprocessing would provide with a best candidate to fetch at each $s_i$ and the algorithm {\bf Matching$_\beta$} would be like Greedy$_{\beta}$, except for the line
\begin{itemize}
\item[-]crobot $\displaystyle\leftarrow \text{arg}\!\min_{q_i\not{in} L} \left\{\text{shortest path from } q_i \text{ to spoint}, R[\text{crobot}]\right\} $ ;
\item[] that should be replaced by
\item[-] crobot$\displaystyle\leftarrow \text{arg}\!\min\left\{q_i|q_is_i\in M_\beta,\: R[\text{crobot}]\right\} $.
\end{itemize} 
Finally, the algorithm {\bf Matching} could return the best of solutions of all {\bf Matching}$_\beta$, having tried all $\beta$s.

Then, the following theorem holds: 
 
\begin{theorem}{\bf Matching} is a $3-$ approximation algorithm for {\sc min-RIPDD}.
\end{theorem}       
{\bf Proof}
%Consider the path of the robot in the optimal that consumes all its energy, in other terms, the path that needs $R^*$ to be walked. Note by $s^*$ the point where this path enters ${\cal P}$. Let $d^*$ be the maximum among distances of robots used in the optimal solution, to their respective entrance in ${\cal P}$ and $d$ be the maximum distance of a robot to its respective selection point, according to $M$.
Let the optimal solution, of value $R^*$, be the solution that fetches $q_1^*,\ldots, q_l^*$ to the selection points $s_1^*\equiv s, s_2^*,\ldots, s_l^*$, respectively. Let $\displaystyle d^*=\max_{1\leq i\leq l}\{d(s_i^*,q_i^*)\}$ and $\displaystyle b^*=\max_{1<i\leq l}\{d_{\cal{P}}(s_i^*,s_{i-1}^*),d_{\cal{P}}(t, s_l^*)\}$. Obviously, we have:

\begin{eqnarray}
R^*&\geq&d^*\label{eq1}\\
R^*&\geq&b^*\label{eq2}
\end{eqnarray}

Consider now the selection points $s_1\equiv s,\ldots, s_{\lfloor\frac{W}{b}\rfloor}$ of the solution provided by {\bf Matching}$_{b}$. As a direct consequence of the Dirichlet's box principle, there is at least one selection point of the optimal in the vertices of each part ${\cal P}_i$ of $\cal{P}$ with endpoints $s_{i}$ and the vertex before $s_{i+1}$ for $1\leq i<\lfloor\frac{W}{b}\rfloor$ (plus the last part with endpoints $s_{\lfloor\frac{W}{b}\rfloor},t$). Let $s^*({\cal P}_i)$ be the selection point of the optimal of the smallest index that belongs to the vertices of ${\cal P}_i$, and let $q^*({\cal P}_i)$ be the agent that the optimal fetches to this selection point. Then   
\begin{eqnarray}
\forall i, 1\leq i\leq \lfloor\frac{W}{b}\rfloor,&d(s^*({\cal P}_i), q^*({\cal P}_i))+b^*\geq d(s_i,q^*({\cal P}_i))\label{eq3}
\end{eqnarray}
(\ref{eq3}) holds also for the $i$ that maximizes $d(s_i,q^*({\cal P}_i))$; but this value is greater than or equal to the maximum value of $M_b$, $d_b$. Thus,
\begin{eqnarray}
d^*+b^*&\geq&d_b\label{eq4}
\end{eqnarray}  
Notice finally, that the value $R$ of the solution returned by the algorithm {\bf Matching} is greater than or equal to the one of {\bf Matching}$_{b^*}$, ie
\begin{eqnarray}
d_b+b^*&\geq&R \label{eq5}
\end{eqnarray} 
So,
\begin{eqnarray}\displaystyle
(\ref{eq1}), (\ref{eq2})\Rightarrow 3R^*\geq d^*+2b^* \geq^{(\ref{eq4})} d_b + b^*\geq^{(\ref{eq5})} R
\end{eqnarray}
$\hfill\qed$

%\begin{figure}[htb]
%\fbox{\parbox[b]{.99\linewidth}{
%\vskip 0.5cm
%\begin{center}
%\includegraphics[width=\columnwidth]{instmatching1.eps}
%\end{center}
%\vskip 0.5cm}}
%\caption{\protect\label{bigr3} $M$ computed for {\bf Matching$_\frac{W}{k}$}. The part of the path in the optimal that equals to $R^*$ is depicted in red.}
%\end{figure}

\section{Discussion}

In this paper we have shown the first - to our knowledge - polynomial time approximation algorithm for the data delivery problem
over a fixed path in a graph performed by mobile agents with identical initial energy budgets and allowing single data hand-overs. Our algorithm generalizes to the case where multiple data hand-overs are allowed, ie whenever an agent that has already carried the data over a part of the path is allowed to hop (following an alternative route) and re-reach the path to carry again the data for a second part of it (given that it still has the energy needed to do so).

Devising exact polynomial algorithms for special graph topologies like the grid is also an interesting open problem; we believe that this variant of data delivery remains difficult for some intuitive special cases of the latter.

\end{document}